\documentclass[reprint,amsmath,amssymb,aps,prb,groupedaddress,nofootinbib,twocolumn,superscriptaddress]{revtex4-2}
\usepackage{graphicx}
\usepackage{amsthm,amssymb,amsmath,braket,mathdots}
\usepackage{bm}
\usepackage[pagebackref=false,pdfnewwindow=true]{hyperref} 
\usepackage{epstopdf,psfrag}
\usepackage{relsize,amsbsy}
\usepackage[export]{adjustbox}

\usepackage{txfonts}

\usepackage{graphicx,xcolor} 
\usepackage{tabularx, booktabs} 

\usepackage{units} 
\usepackage{dsfont} 

\newcommand{\ii}{{\mathrm{i}}} 
\newcommand{\e}{{\mathrm{e}}} 
\renewcommand{\vec}[1]{\bm{#1}} 

\newcommand{\1}{\mathds{1}} 

\newcommand{\ua}{{\uparrow}} 
\newcommand{\da}{{\downarrow}} 

\renewcommand{\eqref}[1]{Eq.~(\ref{#1})}

\usepackage{tikz}
\usetikzlibrary{arrows.meta}

\definecolor{bananayellow}{rgb}{1.0, 0.88, 0.21}
\definecolor{straw}{rgb}{0.32, 0.28, 0.1}

\usepackage[caption=false]{subfig}
\usepackage{changes}
\definechangesauthor[name=Valentin, color=blue]{VL} 
\definechangesauthor[name=Johannes, color=red]{JK}

\begin{document}

\title{Spontaneous Formation of Altermagnetism from Orbital Ordering}
\author{Valentin Leeb}
\affiliation{Technical University of Munich, TUM School of Natural Sciences, Physics Department, TQM, 85748 Garching, Germany}
\affiliation{Munich Center for Quantum Science and Technology (MCQST), Schellingstr. 4, 80799 M{\"u}nchen, Germany}
\author{Alexander Mook} 
\affiliation{Institut f\"ur Physik, Johannes Gutenberg Universit\"at Mainz, D-55099 Mainz, Germany}
\author{Libor {\v{S}}mejkal} 

\affiliation{Institut f\"ur Physik, Johannes Gutenberg Universit\"at Mainz, D-55099 Mainz, Germany}
\affiliation{Institute of Physics, Czech Academy of Sciences, Cukrovarnick\'{a} 10, 162 00 Praha 6 Czech Republic}

\author{Johannes Knolle}
\affiliation{Technical University of Munich, TUM School of Natural Sciences, Physics Department, TQM, 85748 Garching, Germany}
\affiliation{Munich Center for Quantum Science and Technology (MCQST), Schellingstr. 4, 80799 M{\"u}nchen, Germany}
 	\affiliation{\small Blackett Laboratory, Imperial College London, London SW7 2AZ, United Kingdom}
\date{\today}

\begin{abstract}
Altermagnetism has emerged as a third type of collinear magnetism. In contrast to standard ferromagnets and antiferromagnets, 
altermagnets exhibit extra even-parity wave spin order parameters resulting in a spin-splitting of electronic bands in momentum space. In real space, sublattices of opposite spin polarization are anisotropic and related by rotational symmetry. In the hitherto identified altermagnetic candidate materials the anisotropies arise from the local crystallographic symmetry.     
Here, we show that altermagnetism can also form as an interaction-induced electronic instability in a lattice without the crystallographic sublattice  anisotropy. We provide a microscopic example of a two-orbital model showing that the coexistence of staggered antiferromagnetic and orbital order can realize robust altermagnetism. We quantify the spin-splitter conductivity as a key experimental observable and discuss material candidates for the interaction-induced realization of altermagnetism.
\end{abstract}
	
\maketitle

{\it Introduction.--} The interplay of competing orders from strong electronic correlations gives rise to rich phase diagrams of quantum materials,  for example, nematic or antiferromagnetic (AFM) order in the vicinity of superconductivity~\cite{fradkin2015colloquium}. When different orders coexist, qualitatively new behavior can emerge that is not present in the individual phases. For example, superconductivity in materials with long-range magnetism can realize unconventional finite momentum pairing~\cite{fulde1964superconductivity,larkin1965nonuniform}. Here, we show that 
electron correlations can give rise to a phase with coexisting staggered orbital order (OO) and N\'{e}el antiferromagnetism (AFM), which spontaneously realizes a $d$-wave altermagnetic phase with spin-polarized electronic bands and a large transverse spin conductivity.

\begin{figure}[t]
    \centering
    \includegraphics[width=\columnwidth]{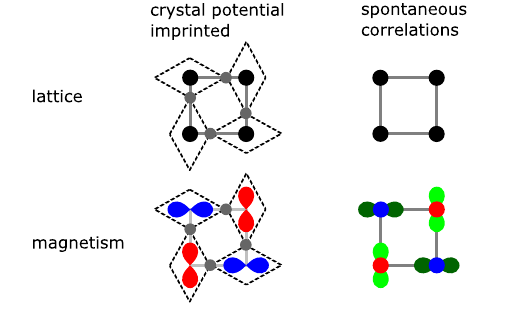}
    \caption{
    Comparison of the established crystal potential-imprinted altermagnetism (left column) and altermagnetism due to spontaneous correlations (right column) on a square lattice. For the former, the crystal structure provides a two site unit cell whose sublattices are related only by rotation but not by inversion symmetry or translation. The symmetry of the lattice, typically due to non-magnetic ions (grey vertices), is imprinted on the electronic 
    density leading to anisotropic spin densities (red, blue) in the magnetically ordered 
    state (lower left). In contrast, for spontaneous altermagnetism the lattice is isotropic and the crucial symmetry lowering happens spontaneously due to a staggered orbital ordering (green).}
    \label{fig1}
\end{figure}


Recently, altermagnetism has been
delimited with the help of spin symmetries as a third type of collinear magnetism~\cite{Smejkal2022altermagnetism}. Similar to standard AFM, an altermagnet displays long-range order with zero net magnetization, e.g. realized by the presence of two sublattices with opposite spin alignment. However, in contrast to usual AFM the N\'{e}el vector is not sufficient for describing an altermagnet because it 
exhibits an extra even-parity wave order parameter~\cite{Smejkal2020,Smejkal2022altermagnetism,Smejkal2022altermagnetism,Smejkal2022perspective,Mazin2022editorial,Mazin2021,Feng2022,Krempasky2023,Bhowal2022,Fernandes2023}. The extra d, g, i-wave spin order takes in momentum space a form of unconventional spin-splitting of the electronic band structure~\cite{Smejkal2022altermagnetism,Smejkal2022perspective,Mazin2022editorial,Krempasky2023} which has recently been confirmed experimentally in photoemission experiments in MnTe~\cite{Krempasky2023,Lee2023}. 
In real space, altermagnets are characterized by anisotropies of spin sublattices, which is best explained with the example of a d-wave 
state on a square lattice, see left column of Fig.~\ref{fig1}. With crystallographic anisotropies, e.g. asymmetric ligands on the bonds, the unit cell has two sites and as a result, in the 
magnetic 
state a flip of all spins is not equivalent to a translation or inversion operation between the spin sublattices but instead they are related by an additional real space rotation~\cite{Smejkal2022altermagnetism}.


So far, research on altermagnets has concentrated on systems where the crystal structure imprints an anisotropic spin density and the two sublattices are globally inequivalent even in the non-magnetic high-temperature phase~\cite{Smejkal2020,Smejkal2022altermagnetism,Smejkal2022perspective,Mazin2021,Mazin2022editorial,Bhowal2022} as in the above example. Below the 
transition temperature it is then the crystal structure which modulates the spin density into an anisotropic shape with opposite spin channels related by crystal rotations or mirrors (possibly non-symmmorphic). Although there is an increasing number of materials of this type, metallic altermagnets are currently rare, and it remains an open question whether altermagnetism can be realized spontaneously via electronic correlations~\cite{wu2007fermi}?
%
Here, we provide an affirmative answer and  demonstrate how altermagnetism can be realized as an OO transition. We concentrate on a minimal two orbital model of transition metal systems with directional $d_{xz/yz}$ orbitals. Crucially, altermagnetism emerges as a \emph{spontaneous} electronic instability in a model whose crystal structure has globally equivalent sublattices, i.e. the lattice does not exhibit the local crystallographic anisotropies. Instead, it is the staggered $(\pi,\pi)$-OO coexisting with $(\pi,\pi)$-AFM, see right column  in Fig.~\ref{fig1}, which leads to the anisotropic spin sublattices, which are again related by rotational symmetry.  
We also show that our proposed mechanism can  generate a strong spin-splitter effect, i.e. a transverse spin-polarized current with spin $d$-wave symmetry~\cite{Gonzalez-Hernandez2021,Smejkal2022GMRTMR} controlled by the OO.

\begin{figure*}
    \centering
    \includegraphics[width=\textwidth]{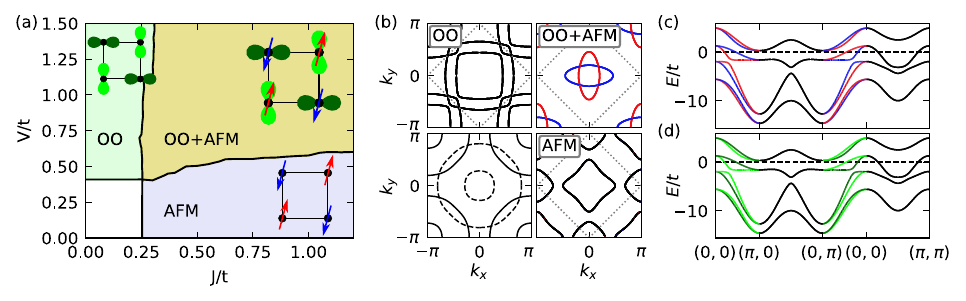}
    \caption{
    Mean-field phase diagram and electronic structure of the model Hamiltonian.
    (a) Zero-temperature mean-field phase diagram for filling $n=5.7216$. There is a trivial phase (white), a phase with $(\pi,\pi)$-antiferromagnetic order (AFM, light blue), a phase with a $(\pi,\pi)$-orbital ordered state (OO, light green), and an altermagnetic phase where AFM and OO are present simultaneously (OO+AFM, ocher). (b) Representative Fermi surfaces of each phase in the crystallographic Brillouin zone. The bands' spin character $\langle \underline s^z\otimes \1 \otimes \1 \rangle$ is indicated by color, with red/blue/black denoting spin-up/spin-down/spin-degenerate. The magnetic Brillouin zone is indicated by gray dotted lines and the dashed contours in the lower left panel (trivial phase) indicate the additional part of the Fermi surface that is present when working in the magnetic $\sqrt{2}\times\sqrt{2}$ unit cell. (c) The electronic band structure along high-symmetry directions in the OO+AFM phase. The bands are both spin-polarized [upper panel, coloring as in (b)] and orbital-polarized (lower panel). Darkgreen (lightgreen) coloring indicates $d_{xz}$ ($d_{yz}$) orbital character $\langle \1 \otimes \1 \otimes \alpha^z\rangle$ and black degeneracy. A $\pi/2$ rotation, which maps the $(0,0)-(0,\pi)$ path onto the $(0,0)-(\pi,0)$ path, also maps oppositely spin- and orbital-polarized bands onto each other. 
    }
    \label{fig2}
\end{figure*}

The example of a d-wave altermagnet has been known for some time~\cite{Smejkal2020,Ahn2019} as the magnetic-analogue of $d$-wave superconductivity~\cite{schofield2009there}. In a long-wavelength momentum space description it can arise as a spin-triplet Pomeranchuk instability of an interacting Fermi liquid, however no realistic candidates have been identified~\cite{wu2007fermi}. Here, we show a realization of the $d$-wave altermagnetic state in a microscopic lattice model of transitional metal systems with non-trivial orbital degrees of freedom, which alludes the real space symmetry properties of altermagnetism. In this context it is interesting to note that the dichotomy between a weak coupling momentum space instability compared to a real space OO instability is similar to the case of Ising nematic order observed in parent compounds of iron-based superconductors~\cite{fernandes2014drives}. There, the breaking of the lattice rotational symmetry can be either understood as a Fermi surface type instability~\cite{fernandes2012preemptive} or a spontaneous OO transition from local Hubbard type interactions~\cite{lee2009ferro,chen2010orbital,lv2009orbital}. Both describe similar physics but while the former approach concentrates on the universal aspects the latter takes into account microscopic details of a given material. Similarly, we show that our microscopic real space description highlights the role of orbital degrees of freedom for realizing  
$d$-wave altermagnetism
from the coexistence of AFM and staggered OO, which allows us to identify a number of possible materials candidates such as perovskites, square pnictides, and vanadates, which we discuss in the outlook section.

{\it Two orbital model.--}
We concentrate on a minimal model of interacting electrons on the square lattice. The full Hamiltonian is given by $H = H_0 + H_J + H_V$. 
The kinetic part $H_0$ consists of two orbitals of the $d_{xz}$ and $d_{yz}$ type described by
\begin{align}
H_0 &= \sum_{\vec{k},s} \vec{\Psi}_{\vec{k}s}^\dagger \begin{pmatrix}
\varepsilon_x(\vec{k}) & \varepsilon_{xy}(\vec{k})\\
\varepsilon_{xy}(\vec{k}) & \varepsilon_y(\vec{k}) 
\end{pmatrix}
\vec{\Psi}_{\vec{k}s}
\label{eq:H0}
\end{align}
with
\begin{align}
\varepsilon_x(\vec{k}) &= -2t_1 \cos k_x - 2t_2 \cos k_y- 4t_3 \cos k_x \cos k_y, \\
\varepsilon_y(\vec{k}) &= -2t_2 \cos k_x - 2t_1 \cos k_y- 4t_3 \cos k_x \cos k_y ,\\
\varepsilon_{xy}(\vec{k}) &= - 4t_4 \sin k_x \sin k_y.
\end{align}
and the components $\Psi_{\vec{k} \alpha s} = 1/\sqrt{N} \sum_j \e^{\ii \vec{k} \cdot \vec{r}_j}\Psi_{j \alpha s}$ of the vector $\vec{\Psi}_{\vec{k}}$ annihilate an electron with momentum $\vec{k}$ and spin $s$ in orbital $\alpha$. $N$ is the number of sites and $\vec{r}_j$ the coordinate of the $j$th unit cell. For concreteness, we fix $t_1=-t$, $t_2=-1.75t$, $t_3=-0.85 t$, $t_4=-0.65 t$ throughout this work.
The anisotropy of the orbitals is imprinted in the hoppings, e.g. $t_1$ quantifies $d_{xz}$ to $d_{xz}$ hopping along the $x$-direction and $d_{yz}$ to $d_{yz}$ hopping along the $y$-direction and $t_2$ $d_{xz}$ ($d_{yz}$) to $d_{xz}$ ($d_{yz}$) hopping along the $y$($x$)-direction. The remaining allowed overlaps are the intraorbital next-nearest neighbor term $t_3$, and $t_4$ as a phase changing interorbital next-nearest neighbor hopping, see the Supplementary Material (SM) for a visualization~\cite{supplement}. The model has been proposed previously as a minimal model for iron-pnictides~\cite{Raghu2008_minimal} but can be adapted to describe any transition metal materials with dominating $d_{xz}$/$d_{yz}$ orbital contributions. 

A natural choice for the interactions would be the local form of the Coulomb repulsion in the Kanamori form~\cite{kanamori1959superexchange} including on-site Hubbard interactions, FM exchange from Hund's coupling, as well as inter-orbital density repulsion and pair hopping~\cite{Raghu2008_minimal,daghofer2010three,dagotto2001colossal}. However, the precise value of the interactions are in general hard to quantify and longer range components of the interaction are neglected. Moreover, many different competing phases can be induced by the interactions depending on the precise form of the Fermi surface. To avoid complications from other competing states, and as a proof of principle, we chose to concentrate on the following effective  interactions 
\begin{align}
H_J = J \sum_{\langle i j \rangle} \vec{S}_i \cdot \vec{S}_j \  \ \text{and} \   \
H_V = V \sum_{\langle i j \rangle} N^z_i N^z_j, 
\end{align}
 with $S^\nu_i = \Psi_{i}^\dag \underline s^\nu \otimes \1 \Psi_{i}$ the total spin at site $i$. The first term is a usual AFM Heisenberg exchange and the second term with  $N^z_i = \Psi_{i}^\dag \1 \otimes \underline{\alpha}^z \Psi_{i} = \sum_s \Psi_{ixs}^\dag \Psi_{ixs}-\Psi_{iys}^\dag \Psi_{iys}$ is an Ising type interaction between the nearest neighbor on-site relative orbital  densities. We denote the Pauli matrix (component $\nu$) for the spin (orbital) subspace by $\underline s^\nu$ ($\underline \alpha^\nu$). We then expect that $H_J$ induces $(\pi,\pi)$-AFM and, crucially, $H_V$ induces $(\pi,\pi)$-OO. 
Because of the absence of spin-orbit coupling, $[H,S^\nu_i]=0$ and the spin remains a good quantum number. Note however, that this is not true for the orbital character, because $t_4$ couples the $d_{xz}$ and $d_{yz}$ orbitals.

{\it Phase diagram.--}
Next, we study the full Hamiltonian in a Hartree--Fock mean-field approximation. In our ansatz, we focus on $(\pi,\pi)$ instabilities and introduce two sublattices $\lambda = A,B$ of even and odd sites. The fermions $\Psi_{\vec{k}\alpha\lambda s}$ of the eight component vector $\vec{\Psi}_{\vec{k}}$ get the new quantum number $\lambda$ and analogously to above $\underline \lambda^\nu$ is the Pauli-$\nu$ matrix acting in the sublattice subspace. 
We can then define the mean fields $\delta m = \sum_i (-1)^{i} \langle S^z_i \rangle/N$, i.e. the order parameter for the staggered AFM, and $\delta n = \sum_i (-1)^{i} \langle N^z_i \rangle/N$, the order parameter for staggered OO. Decoupling the Hamiltonian in the charge channel, we find
\begin{align}
H_J/16J &= -\delta m \sum_{\vec{k}} \vec{\Psi}_{\vec{k}}^\dag \underline s^z  \otimes \underline \lambda^z \otimes \1\vec{\Psi}_{\vec{k}} + 4 \delta m^2, \\
H_V/16V &= -\delta n \sum_{\vec{k}} \vec{\Psi}_{\vec{k}}^\dag \1  \otimes \underline \lambda^z \otimes \underline \alpha^z\vec{\Psi}_{\vec{k}} + 4 \delta n^2 .
\end{align}

The resulting mean-field Hamiltonian $H= \sum_{\vec{k}} \Psi_{\vec{k}}^\dag h(\vec{k}) \Psi_{\vec{k}} + E_0$, see SM~\cite{supplement}, is a non-interacting Hamiltonian, hence the $8\times 8$ Bloch Hamiltonian $ h(\vec{k})$ can be efficiently diagonalized for each momentum to obtain the eigenenergies $\epsilon_m(\vec{k})$ and the Bloch eigenstates $\ket{u_m(\vec{k})}$ with band index $m$. Note that $H$ is block-diagonal in spin, but in contrast to the conventional AFM spin density mean-field solution~\cite{knolle2011multiorbital} the two blocks are explicitly spin-dependent. We have solved the mean-field equations self-consistently for fixed filling $n$ by iteration, i.e. we calculated $\delta m_i$ and $\delta n_i$ from $H(\delta m_{i-1},\delta n_{i-1})$ until convergence, defined by $|\delta m_{i} - \delta m_{i-1}|, |\delta n_{i} - \delta n_{i-1}| < 10^{-3}$,  was reached.

The resulting phase diagram features four different metallic phases, see Fig.~\ref{fig2}~(a). For dominating Heisenberg exchange, i.e. $J/t$ larger than roughly $0.25$ and small $V$, AFM order develops as expected. Conversely, for dominating orbital repulsion between nearest neighbor sites, i.e. $V/t$ larger than $0.5$ and small $J$, OO develops. Crucially, we also find a large coexistence regime (OO+AFM). Importantly, because of the staggered OO the two sublattices of up and down spins are only connected by a real space rotation taking $d_{xz} \to d_{yz}$, but not by translation or inversion as is the case in the pure AFM phase.

The spontaneous symmetry breaking in the OO+AFM phase has important consequences for the electronic structure. Figures~\ref{fig2}~(b,c) show that only in the coexistence OO+AFM phase the spin degeneracy of the bands is removed, as is evidenced by a spin-split Fermi surface (red for spin up and blue for spin down). As expected for the altermagnetic phase, the spin polarized bands are related by $\pi/2$ rotations in momentum space. As an interesting side note, we also find that within our two-orbital only model, the bands are also perfectly orbital polarized as shown in the lower panel of Fig.~\ref{fig2}~(c).

{\it Spin conductivity.--}
Next, we study the unique experimental signatures of the OO+AFM coexistence phase. One of the key features of $d$-wave altermagnets is the  appearance of a
longitudinal spin conductivity without magnetization and a spin-splitter effect \cite{Gonzalez-Hernandez2021}.
The anisotropic spin-polarized Fermi surfaces respond to electric fields by generating characteristic spin currents, see Fig.~\ref{fig2}~(b). When the field is applied in the $[100]$ direction ($x$ direction), the spin-up polarized Fermi surface contributes stronger to transport than the spin-down polarized Fermi surface. As a result, the spin-polarized currents $\vec{j}_\uparrow$ and $\vec{j}_\downarrow$ per Fermi surface are unequal aligned and there is a net spin-polarized current in direction of the electric field.

\begin{figure}
    \centering
    \includegraphics[width=\columnwidth]{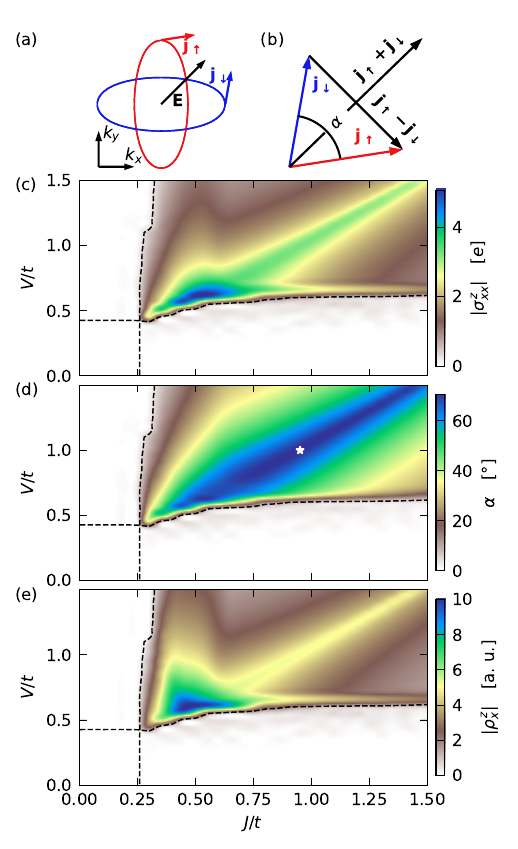}
    \caption{Nonrelativistic spin-polarized currents in the OO+AFM altermagnetic phase. (a) The spin-split Fermi surface in the altermagentic phase leads to a finite spin conductivity $\sigma(\underline s^z) = \sigma^z$, because the spin polarized Fermi surfaces respond with different currents $\vec{j}_\ua, \vec{j}_\da$ to an external electric field $\vec{E}$. The shown Fermi surface has the maximal spin splitter angle. (b) The spin splitter angle $\alpha$ is the angle between the spin currents. Colorplots of the longitudinal component of the spin conductivity $\sigma_{xx}^z = -\sigma_{yy}^z$ (c), the spin splitter angle $\alpha$ (d) ($\alpha_\text{max}$ indicated by star) and the directed spin density of states $\rho^z_{x} = -\rho^z_{y}$ (e). The dashed lines indicate the phase boundaries as extracted in Fig.~\ref{fig2}~(a).}
    \label{fig3}
\end{figure}

To quantify the spin current strength, we evaluate the conductivity along the crystal axis, given in its most general form by the Kubo formula~\cite{Gonzalez-Hernandez2021}
\begin{align}
    \sigma_{bc}(O^a) = -\frac{e \pi}{N} \sum_{\vec{k},n,m} & A_n(\vec{k},\omega) \bra{u_n(\vec{k})} J_b(O^a,\vec{k}) \ket{u_m(\vec{k})} \nonumber\\ \times&  A_m(\vec{k},\omega) \bra{u_m(\vec{k})} v_{c}(\vec{k}) \ket{u_n(\vec{k})}.
    \label{eq:spin_cond}
\end{align}
Here, $A_n(\vec{k},\omega) = -\frac{1}{\pi} \frac{\Gamma}{(\omega-\epsilon_n(\vec{k}))^2+ \Gamma^2}$ is the band-resolved spectral function with a positive infinitesimal broadening $\Gamma$, $v_c = \partial  h(\vec{k})/ \partial k_c$ is the velocity operator and $J_b(O^a) = \frac{1}{2} \{O^a, v_b\}$ the current operator. The summation extends over all eight bands $n,m$ and all momenta $\vec{k}$ of the magnetic Brillouin zone. \eqref{eq:spin_cond} captures charge $\sigma^0 = \sigma(\1)$, spin $\sigma^z = \sigma(\underline s^z)$ and orbital conductivity $\sigma(L^z)$ by adapting the current operator to $J_b(\1) = v_b$, $J_b(\underline s^z) = \underline s^z \otimes \1 \otimes \1 v_b $ or $J_b(L^z) = \frac{1}{2} \{\1 \otimes \1\otimes L^z, v_b\}$ respectively.
We show the magnitude of the respective spin-conductivity tensor element $\sigma_{xx}^z$ in Fig.~\ref{fig3}~(c). The transversal component $\sigma_{xy}^z = 0$ vanishes.

An electric field applied in the $[110]$ direction induces non-parallel spin-currents enclosing the spin splitter angle $\alpha$, see Fig.~\ref{fig3}~(a) and (b). The spin splitter angle $\tan (\alpha/2) = \left|\sigma_{xx}^z/\sigma_{xx}^0 \right |$ quantifies the strength of spin transport compared to standard charge transport and takes its theoretical maximal value of $90^\circ$ for strongly anisotropic elongated Fermi pockets. As shown in Fig.~\ref{fig3}~(d), we find a maximal spin splitter angle $\alpha_\text{max}=70.3^{\circ}$ around $J=V\approx t$ for the Fermi surface shown in Fig.~\ref{fig3}~(a).

Crucially, $\sigma^z$ and $\alpha$ are only non-zero in the altermagnetic phase as evidenced by the spin split Fermi surface. They take the maximum values in the bulk of the OO+AFM phase where $V \approx J$. This shows that for large spin splitting it is more relevant that $\delta m$ and $\delta n$ are of the same size than their individual magnitudes. Unexpectedly, the spin conductivity $\sigma^z$ peaks closely to the AFM phase transition  throughout the entire phase diagram, see Fig.~\ref{figA1} in the SM.

The spin conductivity \eqref{eq:spin_cond} is a direct result of the altermagnetic spin splitting. We can alternatively  quantify the latter by studying a simpler more intuitive quantity, the spin density of states weighted with the Fermi velocity
\begin{align}
    \rho^z_b = \frac{1}{N} \sum_{\vec{k},n} A_n(\vec{k}) \left|\partial_b \epsilon_n(\vec{k})\right| \bra{u_n(\vec{k})} \underline s^z \otimes \1 \otimes \1 \ket{u_n(\vec{k})}.
\end{align}
It is numerically cheaper to compute and behaves analogously to the spin conductivity, see Fig.~\ref{fig3}~(e). Although the spin density of states weighted with the Fermi velocity is not a direct physical observable it can be used as a simple measure to quantify spin splitting for metallic altermagnets.

Deep in the AFM+OO phase the bands are also strongly orbital polarized, see Fig.~\ref{fig2}~(d), resembling the spin character. Therefore one might expect that our findings for the spin conductivity carry over to the orbital conductivity $\sigma(L^z)$. 
In order to evaluate $\sigma(L^z)$, we identify $x$ ($y$) with the atomic orbitals $d_{xz}$ ($d_{yz}$), and from \eqref{eq:spin_cond} use the orbital current operator $J_b(L^z) = \frac{1}{2} \{\1\otimes  \1 \otimes \underline \alpha^y, v_b\}$. It turns out that the orbital conductivity vanishes exactly $\sigma(L^z)=0$. The reason is that the bands are orbital polarized but  only the superpositions $d_{xz} \pm \ii d_{yz}$ are eigenstates of the angular momentum operator $L^z$. This is an artifact of our minimal two-orbital model and in the future it will be interesting to explore the orbital conductivity taking into account the full $d$-orbital manifold of states.

{\it Discussion and Outlook.--}
We have shown that the spontaneous lattice symmetry breaking from OO in conjunction with basic N\'{e}el AFM can give rise to an altermagnetic phase with strongly spin polarized bands. In contrast to existing proposals to look for altermagnets in materials with local crystallographic sublattice anisotropies~\cite{Smejkal2022perspective} our proposed mechanism of interaction-induced OO considerably broadens the range of materials candidates. 
Staggered OO has been experimentally observed on the surface of CeCoIn$_5$~\cite{kim2017atomic} or famously in the perovskite-type transition-metal oxides like LaMnO$_3$~\cite{murakami1998resonant,murakami1998direct,mizokawa1999interplay}.   Another promising material platform are Fe-based square lattices. In fact, recently it was shown that checkerboard AFM order in FeSe subjected to an electric field can generate an altermagnetic state~\cite{Mazin2023a} and the same could be achieved by the presence of staggered OO.

A challenge is that most materials with staggered OO, e.g. C-type, display spin-ordering of a different type, e.g. G-type AFM or FM~\cite{solovyev2006lattice}. In fact, the tendency of staggered OO to show FM is in accordance with the well-known Goodenough-Kanamori rules~\cite{goodenough1963magnetism}, but in certain cases these phenomenological rules may be violated~\cite{oles2006spin,geertsma1996influence,khaliullin2005orbital}. Perhaps the most promising candidate materials in this context are cubic vanadates~\cite{khaliullin2001spin} which have been shown to display staggerd OO coexisting with AFMs~\cite{miyasaka2003spin,ulrich2003magnetic}. The precise form of the OO pattern appears even tuneable via thin film strain engineering~\cite{meley2021strain} which opens the possibility for realizing the required coexistence patterns for an altermagnetic phase. Hence, in the future it will be important to explore microscopic scenarios for realistic material platforms.

Beyond the strong coupling analysis, it would also be worthwhile to start with a weak-coupling theory and microscopic interaction parameters to analyze when the AFM and OO susceptibilities of the metallic phase diverge simultaneously. In that context, we note that our coexistence phase of AFM and OO has already been discussed as a potential instability of more generic models, e.g. see Fig.~2~(f) of Ref.~\cite{Daghofer2010_three}, but a systematic study how to stabilize these is missing. Similarly, studies of the dynamical response functions in the coexistence regime will be important for making contact to scattering experiments. 

In conclusion, we have shown how electronic correlations can lead to the spontaneous formation of altermagnetic phases due to OO. As this considerably broadens to range of candidate materials we hope that our work is a stepping stone for realizing new altermagnets. 

\section*{Data and code availability}
Code and data related to this paper are available on Zenodo \cite{code} from the authors
upon reasonable request.

\begin{acknowledgments}
We thank M. Knap for helpful discussions and related collaborations. J.~K. especially thanks K. Wohlfeld for valuable  discussions and for pointing out the connection to Goodenough-Kanamori rules as well as vanadates. V.~L. acknowledges support from the Studienstiftung des deutschen Volkes. J.~K. acknowledges support from the Imperial-TUM flagship partnership. The research is part of the Munich Quantum Valley, which is supported by the Bavarian state government with funds from the Hightech Agenda Bayern Plus. This work is funded in part by the Deutsche Forschungsgemeinschaft (DFG, German
Research Foundation) - Project No.~504261060.
L.~S. acknowledges support from the Johannes Gutenberg-Universität Mainz TopDyn initiative (project ALTERSEED), and funding from Deutsche Forschungsgemeinschaft (DFG) grant no. TRR 288 - 422213477 (Project B05). 
\end{acknowledgments}

\bibliography{bib}

\newpage

\appendix
\begin{widetext}
\section*{Supplementary Material}





\section{Model and Hamiltonian}
We show the hopping structure of the non-interacting part of the Hamiltonian $H_0$ in Fig.~\ref{figA0}. Fourier transformation leads to \eqref{eq:H0} \cite{Raghu2008_minimal}.

 After mean-field decoupling we obtain $H = \sum_{\vec{k}} \vec{\Psi}_{\vec{k}}^\dag h(\vec{k}) \vec{\Psi}_{\vec{k}} + E_0$ with 
\begin{align}
h(\vec{k}) &= \1 \otimes T(\vec{k}) - 16 J\delta m (\tau^z \otimes \tau^z \otimes \1)  - 16 V\delta n (\tau^z \otimes \1 \otimes \tau^z)
\end{align}
and $T(\vec{k})$ is a $4\times4$ matrix with
\begin{align}
    T_{11}(\vec{k}) &= T_{22}(\vec{k}) = T_{33}(\vec{k}) = T_{44}(\vec{k}) =  -2 t_3 (\cos(k_x+k_y)+\cos(k_x-k_y))-\mu \nonumber\\
    T_{12}(\vec{k}) &= T_{21}(\vec{k}) = T_{43}(\vec{k}) = T_{34}(\vec{k})= -2t_4(\cos(k_x+k_y)-\cos(k_x-k_y)) \nonumber\\
    T_{13}(\vec{k}) &= T_{31}(\vec{k})^* = -t_1 \left(1+\e^{2\ii k_x}\right) - 2 t_2 \e^{\ii k_x}\cos(k_y) \nonumber \\
    T_{24}(\vec{k}) &= T_{42}(\vec{k})^* = -t_2 \left(1+\e^{2\ii k_x}\right) - 2 t_1 \e^{\ii k_x}\cos(k_y)
\end{align}
in the basis $(\Psi_{\vec{k} x A \uparrow},\Psi_{\vec{k} y A \uparrow},\Psi_{\vec{k} x B \uparrow},\Psi_{\vec{k} y B \uparrow},\Psi_{\vec{k} x A \downarrow},\Psi_{\vec{k} y A \downarrow},\Psi_{\vec{k} x B \downarrow},\Psi_{\vec{k} y B \downarrow})$ and the energy constant $E_0 = 64 J \delta m^2 + 64 V \delta n^2$. Note that $\tau^z$ is the Pauli-$z$ matrix.

\begin{figure}
    \centering
    \includegraphics[scale=1]{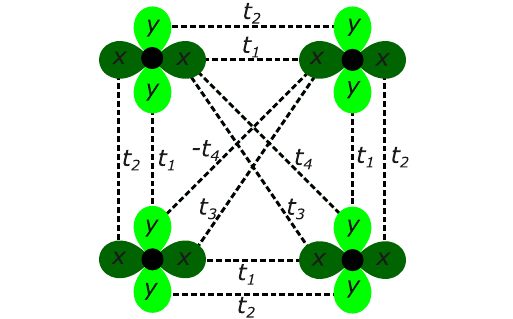}
    \caption{Visualization of Hamiltonian $H_0$. The present hopping terms are shown as dashed lines between the $x$ and $y$ orbital.}
    \label{figA0}
\end{figure}

\section{Full phase diagram}
In principle our model has at least four different tunable parameters, i.e. the interactions $J,V$, temperature and the filling $n$. Based on the well understood effect of the interactions, we do not expect to find any additional instabilities in the phase diagram. For completeness, we show the phase diagram and the spin conductivity $\sigma^z$ for various fillings $n$ and interaction strengths, but fixed interaction ratio $J=V$ in Fig.~\ref{figA1}. 

We observe that the presence of OO and AFM order as well as their coexistence is generic for any filling. OO is dominant at half filling whereas AFM order features its minimal critical interaction around quarter and three quarter filling. Large regions of the phase diagram are again dominated by the coexistence of OO and AFM, i.e. the altermagnetic phase. At exactly half filling an insulator forms which is expected to become extended for more generic Hubbard-like interactions. 

For finite temperatures we found rather conventional behavior, i.e. the two different critical interaction strengths decrease with increasing temperature, leading to two independent critical temperatures.

\begin{figure*}
    \centering
    \includegraphics[scale=1]{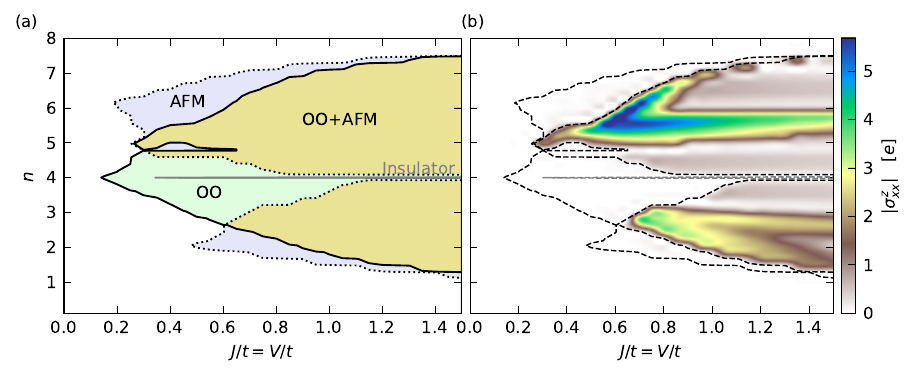}
    \caption{(a) Phase diagram for various fillings $n$ and $V=J$. At half filling an OO insulator emerges. The altermagnetic phase, where OO and AFM order coexist, is always a metal. (b) Colorplot of the spin conductivity $\sigma^z_{xx}$. The phase boundaries of (a) are shown as dashed lines.}
    \label{figA1}
\end{figure*}

\end{widetext}

\end{document}